\documentstyle[conf_iap,flushrt,epsf,rotate,11pt]{article}

\def\rn{}
\def\nn#1 #2{#1 #2.}				
\def\nnn#1 #2 #3{#1 #2. #3.}			
\def\nnnn#1 #2 #3 #4{#1 #2. #3. #4.}		
\def\nnnnn#1 #2 #3 #4 #5{#1 #2. #3. #4. #5.}	
\def\dualand{ and\hbox{ }}				
\def\multiand{ \&~}				
\def\rf#1;#2;#3;#4;#5 {{\frenchspacing\par\rn#1, #2, #3 #4, #5 \par}}
\def\rfbook#1;#2;#3;#4;#5 {{\frenchspacing\par\rn#1, {\it #3} (#5, #4, #2).\par}}
\def\rfprep#1;#2;#3 {{\par\frenchspacing\rn#1 #2, #3\par}}

\def\expec#1{\langle#1\rangle}

\def\etal{{\frenchspacing\it et al.}}
\def\ie{{\frenchspacing\it i.e.}}
\def\eg{{\frenchspacing\it e.g.}}

\def\beq#1{\begin{equation}\label{#1}}
\def\eeq{\end{equation}}
\def\beqa#1{\begin{eqnarray}\label{#1}}
\def\eeqa{\end{eqnarray}}
\def\eq#1{equation~(\ref{#1})}

\def\spose#1{\hbox to 0pt{#1\hss}}
\def\simlt{\mathrel{\spose{\lower 3pt\hbox{$\mathchar"218$}}
     \raise 2.0pt\hbox{$\mathchar"13C$}}}
\def\simgt{\mathrel{\spose{\lower 3pt\hbox{$\mathchar"218$}}
     \raise 2.0pt\hbox{$\mathchar"13E$}}}
\def\simpropto{\mathrel{\spose{\lower 3pt\hbox{$\mathchar"218$}}
     \raise 2.0pt\hbox{$\propto$}}}

\def\ed{\end{document}}

\def\dt{\delta}
\def\dl{g}

\def\k{{\bf k}}
\def\r{{\bf r}}

\def\x{{\bf x}}

\def\kh{\widehat{\bf k}}
\def\rh{\widehat{\bf r}}
\def\xh{\widehat{\x}}
\def\Px{P_\times}
\def\Om{\Omega_m}

\def\P{{\bf P}}

\def\Pobs{P_{obs}}

\def\figone{
\makebox{
\smallskip
\noindent
\parbox[l]{3.5truecm}{\footnotesize
\flushleft
{\bf Figure 1.}
The evolution of bias (A) and correlation
(B) is shown for 15 models that at 
Einstein-de Sitter redshift $z=5$ have 
bias $b=0.5$, 1, and 2 and correlation coefficients $r=0$, 0.25,
0.5, 0.75, and 1. In the top panel, $r$ increases upward 
in each quintuplet of lines. In the bottom panel, 
$b_0$ increases downward in each triplet of lines. 
}
\parbox[r]{8.0truecm}{
\epsfxsize=8.0truecm\epsfbox{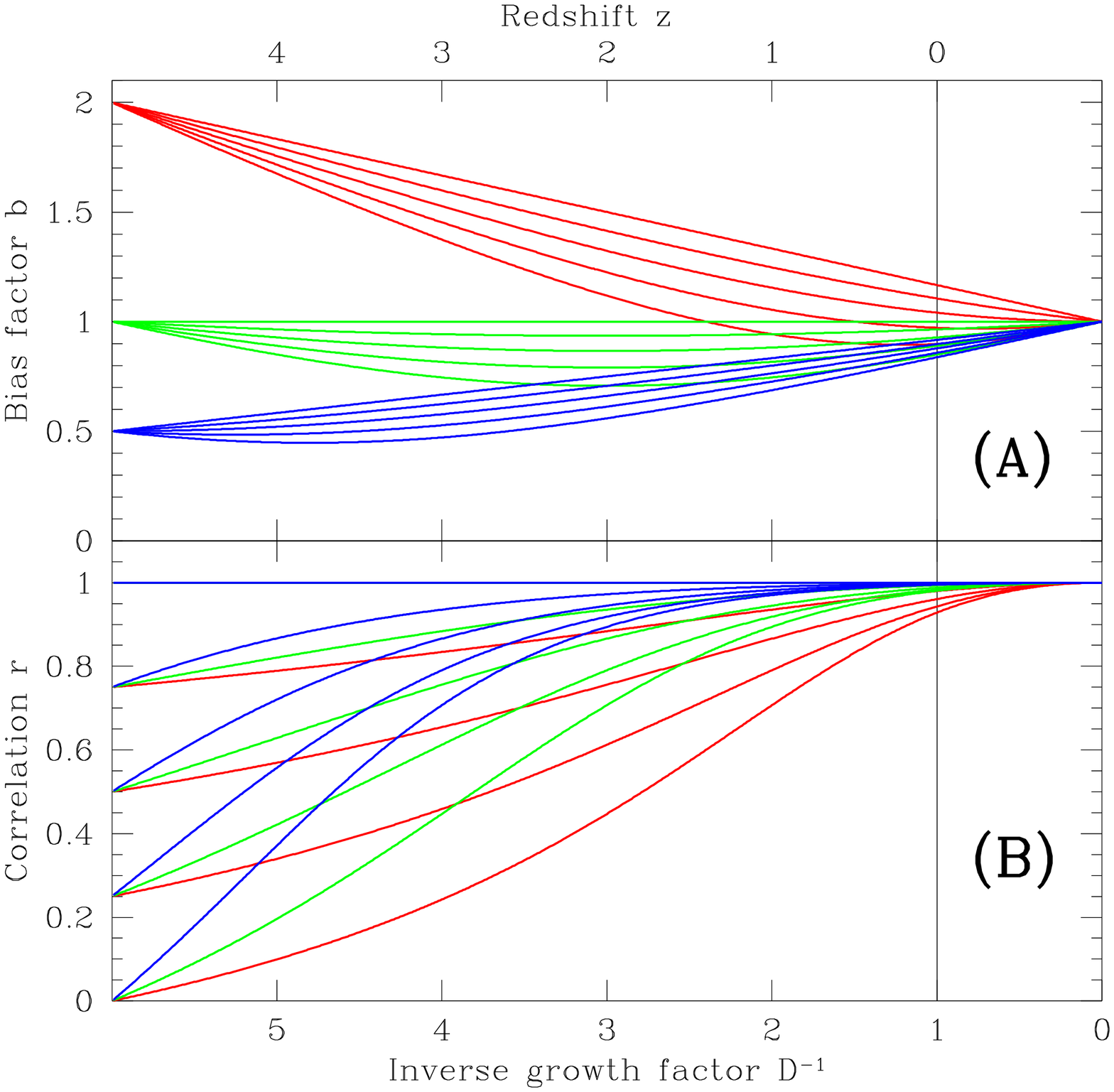}
}
}
}

\begin{document}

\heading{Bias and beyond}
\vskip-3cm
\noindent
To appear in {\it Proceedings of the 14$^{\rm th}$ IAP 
Colloquium: \\
Wide Field Surveys in Cosmology}, 
Eds. S.~Colombi \& Y.~Mellier, \\
Editions Frontieres (1998)
\vskip1.5cm
\par\medskip\noindent

\author {Max Tegmark$^{1,2}$}
 \address{Institute for Advanced Study, Olden Lane, Princeton, NJ 08540}
 \address{Hubble Fellow}

\begin{abstract}
It is becoming increasingly clear that galaxy bias is complicated,
with evidence indicating that it 
can be time-dependent, morphology-dependent, scale-dependent,
non-linear and non-deterministic.
We discuss strategies for overcoming these 
complications and performing precision cosmological
tests with upcoming redshift surveys.
\end{abstract}

\section{The Bad News}

The relative distribution of galaxies and mass is of increasing
concern in cosmology since
constraints on cosmological parameters from galaxy 
surveys \cite{galfisher,Goldberg,Hu,Eisenstein}
are only as accurate
as our understanding of bias.
Unfortunately, bias is complicated. 
The commonly used assumption that the matter density fluctuations 
$\delta(\r)$ and the galaxy number density fluctuations $g(\r)$
obey
\beq{SimpleBiasEq}
g(\r) = b\,\delta(\r)
\eeq
for some constant $b$ (the bias factor) appears to be violated
in a number of ways.
It has been known since the outset \cite{Dressler} that
$b$ must depend on galaxy type. However, there is also evidence that 
it depends on scale (see {\eg} \cite{Mann,Blanton} and references therein) and on time.
Time-evolution of bias is both expected theoretically 
\cite{Fry,bias} and observed \cite{Giavalisco},
with $b$ generally dropping towards unity from high values early on.
This is due to two separate effects: (1) Early galaxies are more biased at
birth since they tend to be associated with rarer peaks.
(2) Old galaxies gradually debias as they are gravitationally
pulled into the same overdensities as the dark matter (see Figure 1). 
Moreover, on small scales where non-linear dynamics is important, one expects nonlinear
corrections to \eq{SimpleBiasEq}. Finally, 
there are good reasons to believe that there is {\it no} deterministic 
relation that can replace \eq{SimpleBiasEq}, but that bias is inherently 
somewhat stochastic \cite{Dekel,Pen,bias,Blanton}.

\section{The Good News}

Does this mean that galaxy redshift surveys will contribute nothing
to the quest for accurate measurements of cosmological parameters?
Not necessarily. Below we will describe an approach that 
may be able to take us beyond the murky marshes of bias
and let us do precision cosmology without assuming any bias relation in place 
of \eq{SimpleBiasEq}.

Dekel and Lahav  
have proposed a framework termed
{\it stochastic biasing} \cite{Dekel}, which drops the assumption that
the galaxy fluctuations $g(\r)$ is uniquely determined by the 
matter fluctuations $\delta(\r)$.
Instead, $\dl$ is modeled as a function of $\dt$ 
plus a random term.
Restricting attention to second moments, 
all the information about stochasticity is contained in a single new
function $r(k)$ \cite{Pen,bias}. 
Grouping the fluctuations into a two-dimensional vector
\beq{xDefEq}
\x\equiv\left({\dt\atop\dl}\right)
\eeq
and assuming nothing except translational invariance,
its Fourier transform $\xh(\k)\equiv\int e^{-i\k\cdot\r}\x(\r)d^3r$ obeys 
\beq{MatrixPowerEq}
\expec{\xh(\k)\xh(\k')^\dagger}=(2\pi)^3\delta^D(\k-\k')
\left(\begin{tabular}{cc}
$P(\k )$ & $\Px(\k )$ \\
$\Px (\k )$ & $P_g(\k )$
\end{tabular}\right)
\eeq
for some $2\times 2$ power spectrum matrix that we will denote $\P(\k)$.
Here $P$ is the conventional power spectrum of the mass distribution, 
$P_g$ is the power spectrum of the galaxies, and $\Px$ is the cross spectrum. 
It is convenient to rewrite this covariance matrix as 
\beq{PdefEq}
\P(\k ) =
P(\k)\left(\begin{tabular}{cc}
$1$ & $b(\k)r(\k)$ \\ $b(\k)r(\k)$ & $b(\k)^2$ \end{tabular}\right)
\eeq
where $b\equiv(P_g/P)^{1/2}$ is the bias factor
(the ratio of galaxy and total fluctuations)
and the new function $r\equiv\Px/(P P_g)^{1/2}$ is the dimensionless 
correlation coefficient between galaxies and matter.
Note that both $b$ and $r$ generally depend on scale $k$.
The Schwarz inequality shows that the special case 
$r=1$ implies the simple deterministic \eq{SimpleBiasEq},
and the converse is of course true as well.

Since the function $r(k)$ is a cosmologically important quantity,
it has received much recent attention. 
Pen has shown how it can be measured using 
redshift space distortions and nonlinear effects \cite{Pen}, and 
it has recently been computed for a 
number of theoretical models \cite{Scherrer}.
Its time-evolution has been derived in the linear regime \cite{bias}
and generalized to the perturbatively non-linear case \cite{Taruya}.
For a fixed galaxy population in an
in an Einstein-de Sitter universe ($\Omega_m=1$ and 
$\Omega_\Lambda=0$, as in the standard cold dark matter
model), the relation between the values $(b,r)$ at redshift $z$ 
and the present values $(b_0,r_0)$ reduces to \cite{bias}
\beqa{CoolEq3}
b     &=&[z^2 - 2z(1+z)b_0 r_0 + (1+z)^2 b_0^2]^{1/2},\label{bEq3}\\
r     &=&[(1+z)b_0 r_0 - z]/b\label{req3},
\eeqa
plotted in Figure 1.
Both $b$ and $r$ have now been measured 
using hydrodynamic simulations \cite{Blanton}, and 
other groups are performing related studies.

\noindent\figone

Why is this useful? How can one learn anything about the matter density 
without making assumptions about how it relates to the
galaxy density? By assuming that the {\it velocities}
are the same for galaxies as for other matter on large scales, 
\ie, that they are both caused by gravity.
This means that the characteristic clustering anisotropies
known as redshift-space distortions \cite{Kaiser,Hamilton}
will appear even if galaxies are completely uncorrelated.
Within the stochastic biasing framework, 
Kaiser's classic result (which assumes $r=1$) gets generalized to
\cite{Pen,Dekel}
\beq{RightKaiserEq}
\Pobs(\k) = \left[b(k)^2 + 2\Om^{0.6} b(k) r(k) x + \Om^{1.2} x^2\right]P(\k),
\eeq
where $x\equiv (\rh\cdot\kh)^2$ and $\rh$ is the line of sight
direction (for the special case of a volume-limited distant survey).
The right-hand side is a quadratic polynomial in $x$, 
so with a large data set such as the 2-degree Field Survey (2dF)
or the Sloan Digital Sky Survey (SDSS), one can accurately measure
the three coefficients.
For each scale $k$, \eq{RightKaiserEq} thus gives us three independent
equations.
If the matter density $\Omega_m$ can be accurately
measured by other means
(using say upcoming CMB experiments, supernova 1a surveys 
and cluster abundance evolution),
then these three equations are readily solved 
for the three unknowns $P(k)$, $b(k)$ and $r(k)$.
The last two would then provide valuable 
insights about galalaxy formation, whereas 
$P(k)$ would give us the long sought for power spectrum of the
underlying matter distribution which could, among other things,  
provide precision constraints on
neutrino masses \cite{Hu} and the Hubble constant \cite{Eisenstein}.

\smallskip
The author wishes to thank John Bahcall, Michael Blanton,
Ben Bromley, Avishai Dekel, Daniel Eisenstein, Wayne Hu, 
Ofer Lahav, Jerry Ostriker, Jim Peebles, Ue-Li Pen, 
Martin Rees and Michael Strauss
for useful discussions.

This work was supported by NASA grant NAG5-6034 and Hubble Fellowship 
HF-01084.01$-$96A from by STScI, operated by AURA, Inc. under 
NASA contract NAS5-26555.


\end{document}